\documentclass[a4paper]{article}
\usepackage{subfigure}
\usepackage{hyperref}
\usepackage{multirow}
\usepackage{INTERSPEECH2021}

\title{Rich Prosody Diversity Modelling with Phone-level Mixture Density Network}
\name{Chenpeng Du and Kai Yu}
\address{
MoE Key Lab of Artificial Intelligence, AI Institute\\
X-LANCE Lab, Department of Computer Science and Engineering\\
Shanghai Jiao Tong University, Shanghai, China}
\email{\{duchenpeng, kai.yu\}@sjtu.edu.cn}

\begin{document}

\maketitle
\begin{abstract}
Generating natural speech with diverse and smooth prosody pattern is a challenging task. Although random sampling with phone-level prosody distribution has been investigated to generate different prosody patterns, the diversity of the generated speech is still very limited and far from what can be achieved by human. This is largely due to the use of uni-modal distribution, such as single Gaussian, in the prior works of phone-level prosody modelling. In this work, we propose a novel approach that models phone-level prosodies with GMM based mixture density network (GMM-MDN). Experiments on the LJSpeech dataset demonstrate that phone-level prosodies can precisely control the synthetic speech and GMM-MDN can generate more natural and smooth prosody pattern than a single Gaussian. Subjective evaluations further show that the proposed approach not only achieves better naturalness, but also significantly improves the prosody diversity in synthetic speech without the need of manual control.
\end{abstract}

\section{Introduction}

Neural text-to-speech(TTS) synthesis models with sequence-to-sequence architecture \cite{tacotron,tacotron2,transformertts} can be applied to generate naturally sounding speech. Recently, non-autoregressive TTS models such as FastSpeech \cite{fastspeech} and FastSpeech2 \cite{fastspeech2} are proposed for fast generation speed without frame-by-frame generation. 

Besides the progress of acoustic modelling, prosody modelling is also widely investigated.
Utterance level prosody modelling in TTS is proposed in \cite{ref_speech}, in which an utterance-level prosody embedding is extracted from a reference speech for controlling the prosody of TTS output. \cite{gst} factorizes the prosody embedding with several global style tokens(GST).
Variational auto-encoder(VAE) is used for prosody modelling in \cite{tacotron_vae}, which enables us to sample various prosody embeddings from the standard Gaussian prior in inference stage. 

Utterance-level prosody is hard to precisely control the synthetic speech due to its coarse granularity. Therefore, phone-level prosody modelling is also analyzed in recent works.
\cite{prosody_attention} extracts prosody information from acoustic features and uses an attention module to align them with each phoneme encoding. \cite{phone_level_prosody_amazon} directly models phone-level prosody with a VAE, thus improving the stability compared with \cite{prosody_attention}.
Hierarchical and quantized versions of VAE for phone-level prosody modelling is also investigated in \cite{phone_level_3dim,phone_level_3dim_quantize,japan_hierarchical_vae,lhy_pl_vae}, which improves the interpretability and naturalness in synthetic speech. 

However, most the prior works for phone-level prosody modelling assumes that the distribution of phone-level prosodies is uni-modal distribution, such as a single Gaussian, which is not reasonable enough and hence can only provide limited diversity.
Actually, phone-level prosodies are highly diverse even for the same context, hence it is natural to apply multi-modal distribution. In traditional ASR systems, one of the most dominant techniques is HMM-GMM \cite{gmmasr1,gmmasr2,gmmasr3}, in which the distribution of acoustic features for each HMM state is modeled with a GMM. Similarly, GMM is also used to model acoustic features in traditional statistical parametric speech synthesis(SPSS) \cite{tts_mdn1,tts_mdn2}, thus improving the voice quality.

Inspired by these prior works, we propose a novel approach that models phone-level prosodies with GMM based mixture density network (GMM-MDN) \cite{mdn}.
We use a prosody extractor to extract phone-level prosody embeddings from ground-truth mel-spectrograms and use a prosody predictor as the MDN to predict the GMM distribution of the embeddings. 
In inference stage, the prosody of each phoneme is randomly sampled from the predicted GMM distribution for generating speech with diverse prosodies. 

Our experiments on the LJSpeech \cite{ljspeech} dataset demonstrate that phone-level prosodies can precisely control the synthetic speech and GMM-MDN can generate more natural and smooth prosody pattern than a single Gaussian. The subjective evaluations suggest that our method not only achieves a better naturalness, but also significantly improves the prosody diversity in synthetic speech without the need of manual control.

In the rest of this paper, we first review the MDN in Section \ref{sec:mdn} and introduce the proposed model in Section \ref{sec:pl_modelling}. Section \ref{sec:setup} gives experiments comparison and results analysis, and Section \ref{sec: conclusion} concludes the paper.

\begin{figure*}[htb]
  \centering
  \subfigure[Overal architecture based on FastSpeech2]{
    \includegraphics[height=5.6cm]{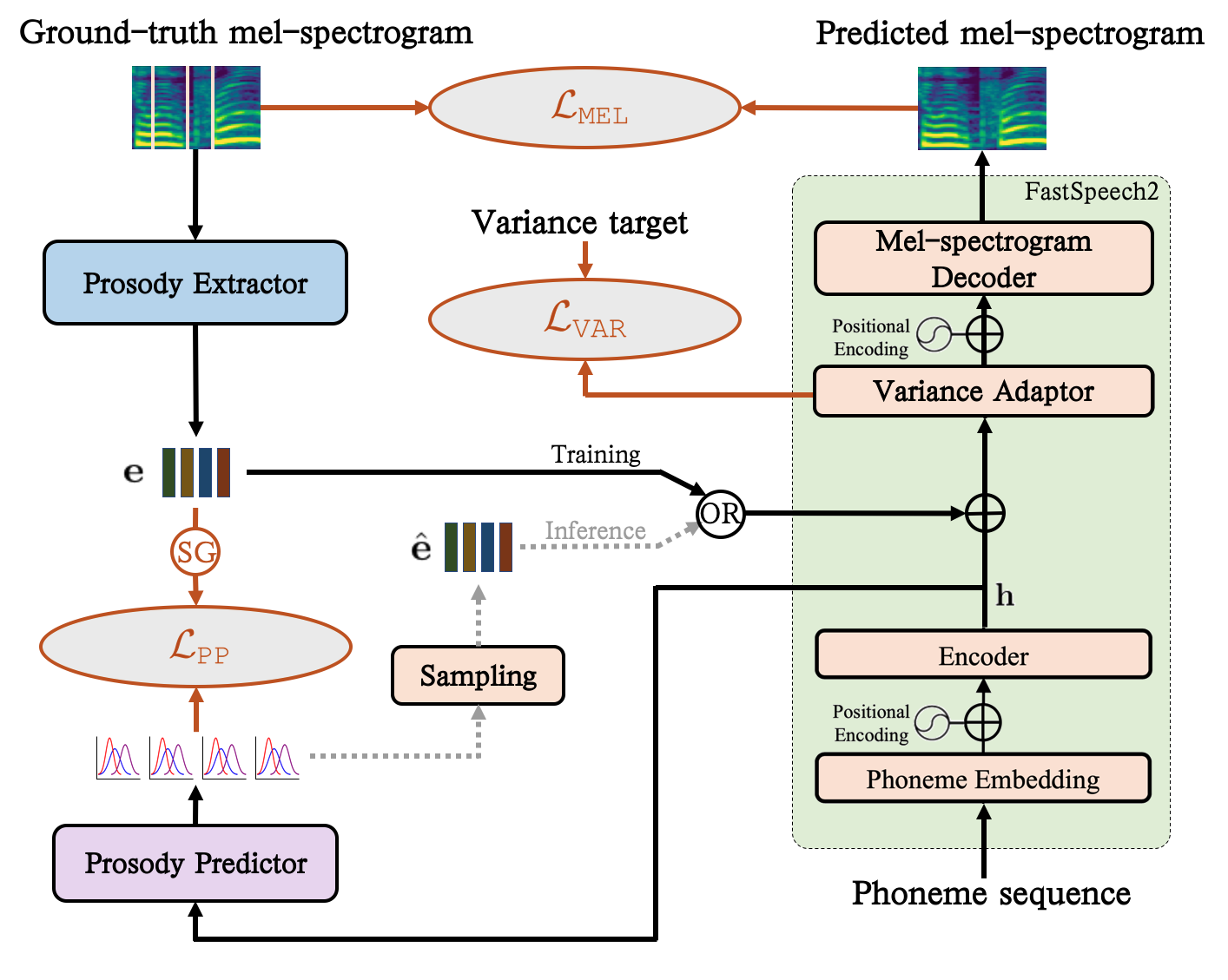}
    \label{overall}
  }
  \subfigure[Prosody extractor]{
    \includegraphics[height=6.5cm]{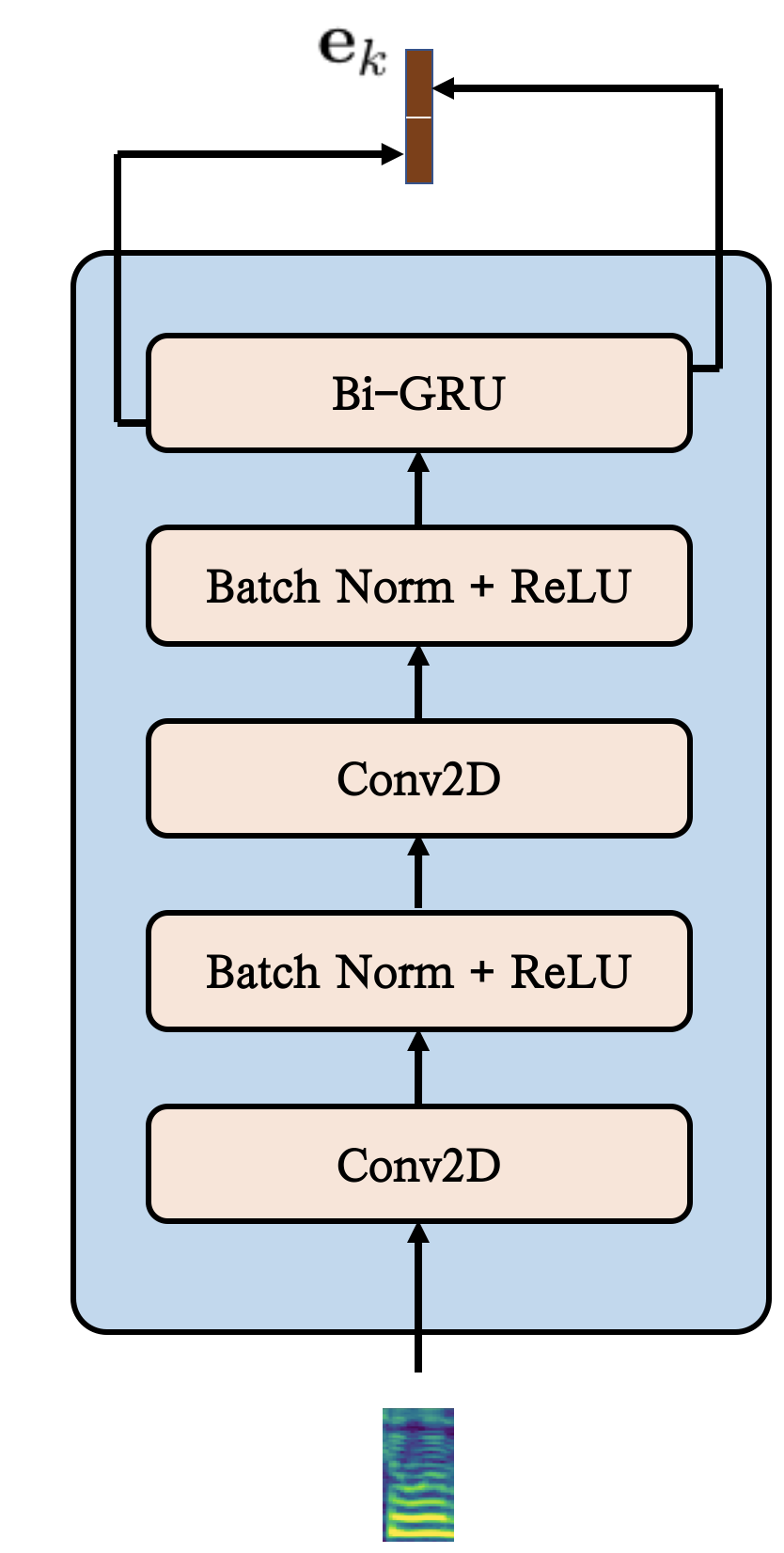}
    \label{extractor}
  }
  \subfigure[Prosody predictor]{
    \includegraphics[height=7.5cm]{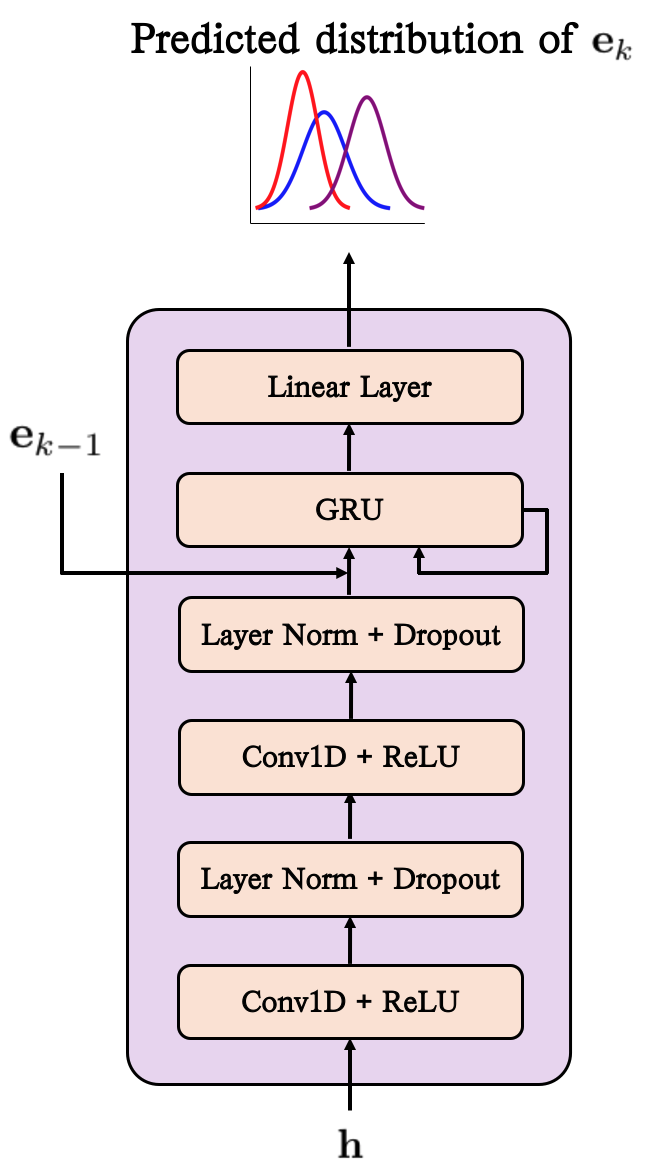}
    \label{predictor}
  }
  \caption{GMM-based phone-level prosody modelling. ``SG'' represents the stop gradient operation. ``OR'' selects the extracted ``ground-truth'' $\mathbf{e}$ in training and the sampled $\hat{\mathbf{e}}$ in inference. We use red lines for loss calculation and dash lines for sampling.}
  \label{fig:single_spk_model}
\end{figure*}

\section{Mixture Density Network}
\label{sec:mdn}
 Mixture density network \cite{mdn} is defined as the combined structure of a neural network and a mixture model. We focus on GMM based MDN in this work to predict the parameters of the GMM distribution, including the means $\mu_{i}$, variances $\sigma_{i}^{2}$, and mixture weights $w_i$. It should be noted that the sum of the mixture weights is constrained to 1, 
which can be achieved by applying a Softmax function, formalized as
\begin{equation}
\begin{aligned}
\label{eq:softmax}
w_{i}=\frac{\exp \left(\alpha_{i}\right)}{\sum_{j=1}^{M} \exp \left(\alpha_{j}\right)}
\end{aligned}
\end{equation}
where $M$ is the number of Gaussian components and $\alpha_{i}$ is the corresponding neural network output. The mean and variance of Gaussian components are presented as
\begin{eqnarray}
 \mu_{i}& = & m_{i} \\\label{eq:mean}
 \sigma_{i}^{2} & = & \exp \left(v_{i}\right)\label{eq:var}
\end{eqnarray}
where $m_{i}$ and $v_{i}$ are the neural network outputs corresponding to the mean and variance of the $i$-th Gaussian component. Equation (\ref{eq:var}) constrains the $\sigma_{i}^{2}$ to be positive.

The criterion for training the MDN is the negative log-likelihood of the observation $\mathbf{y}$ given its input $\mathbf{x}$. Here we can formulate the loss function as
\begin{equation}
\begin{aligned}
\label{eq:mdn_loss}
\mathcal{L}_{\texttt{MDN}} & = -\log p\left(\mathbf{y};\mathbf{x}\right) \\
& = -\log \left(\sum_{i=1}^{M} w_{i} \cdot \mathcal{N}\left(\mathbf{y} ; \mu_{i}, \sigma_{i}^{2}\right)\right)
\end{aligned}
\end{equation}
Therefore, given the input $\mathbf{x}$, the mixture density network is optimized to predict GMM parameters $w_i$, $\mu_i$ and $\sigma_i$ that maximize the likelihood of $\mathbf{y}$.

\section{GMM-MDN for Phone-Level Prosody Modelling}
\label{sec:pl_modelling}

The TTS model in this paper is based on the recent proposed FastSpeech2\cite{fastspeech2}, where the input phoneme sequence is first converted into a hidden state sequence $\mathbf{h}$ by the encoder and then passed through a variance adaptor and a decoder for predicting the output mel-spectrogram. Compared with the original FastSpeech\cite{fastspeech}, FastSpeech2 is optimized to minimize the mean square error(MSE) $\mathcal{L}_\texttt{MEL}$ between the predicted and the ground-truth mel-spectrograms, instead of applying a teacher-student training. Moreover, the duration target is not extracted from the attention map of an autoregressive teacher model, but from the forced alignment of speech and text. Additionally, \cite{fastspeech2} condition the prediction of mel-spectrogram on the variance information such as pitch and energy with a variance adaptor. The adaptor is trained to predict the variance information with an MSE loss $\mathcal{L}_\texttt{VAR}$.

In this work, we propose a novel approach that models phone-level prosodies with a GMM-MDN in the FastSpeech2-based TTS system. Accordingly, we introduce a prosody extractor and a prosody predictor, whose architecture and training strategy are described below.

\subsection{Overall architecture}
The overall architecure of the proposed system is shown in Figure \ref{overall}.
In the training stage, the prosody embeddings 
\begin{equation}
\begin{aligned}
\label{eq:idx_seq}
\mathbf{e}=[\mathbf{e}_1, \mathbf{e}_2, ..., \mathbf{e}_K]
\end{aligned}
\end{equation}
are extracted for all the $K$ phonemes by the prosody extractor from the corresponding mel-spectrogram segment. It is then projected and added to the corresponding hidden state sequence $\mathbf{h}$ in order to better reconstruct the mel-spectrogram.
We use $\mathbf{e}_k$ to represent the prosody embedding for the $k$-th phoneme. The distribution of $\mathbf{e}_k$ is assumed to be a GMM whose parameters are predicted by an MDN. Here, the GMM-MDN is the prosody predictor.
The prosody predictor autoregressively predicts the GMM distributions of the prosody embeddings. In inference stage, we sample the $\hat{\mathbf{e}}_k$ from the predicted distribution for each phoneme, so that we can generate speech with diverse prosodies.

\subsection{Prosody extractor}

The detailed architecture of the prosody extractor is shown in Figure \ref{extractor}. It contains 2 layers of 2D convolution, each followed by a batch normalization layer and a ReLU activation function. A bidirectional GRU is designed after the above modules. The concatenated forward and backward states from the GRU layer is the output of the prosody extractor, which is referred to as the prosody embedding of the phoneme.

\begin{figure*}[t]
\centering
\subfigure[LJSpeech training set]{
\includegraphics[width=0.4\linewidth]{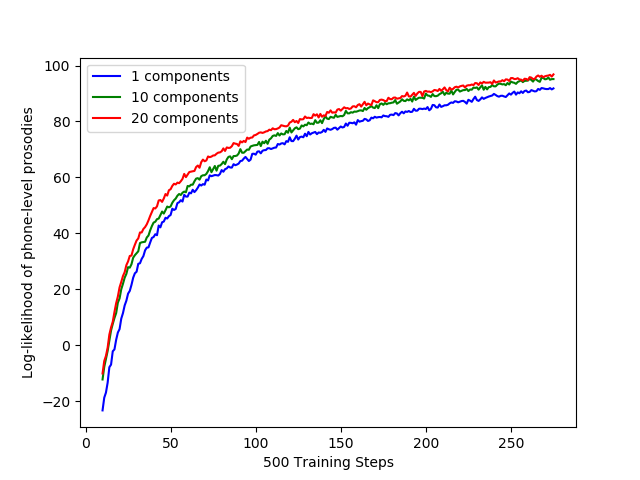}
\label{fig:loglike_ljspeech_train}
}
\subfigure[LJSpeech test set]{
\includegraphics[width=0.4\linewidth]{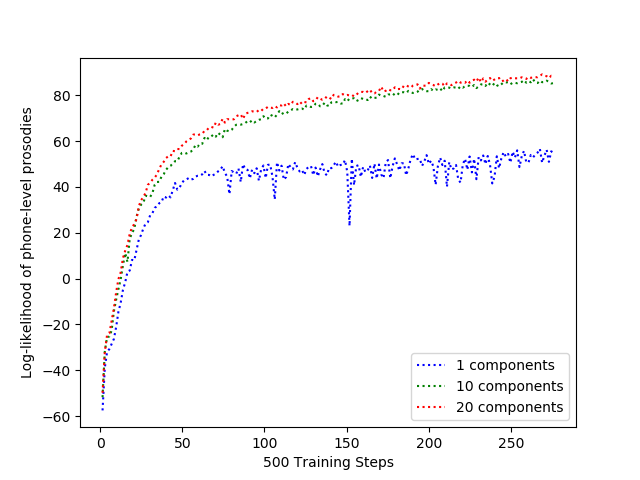}
\label{fig:loglike_ljspeech_valid}
}
\caption{Log-likelihood curves of the extracted phone-level prosodies with different numbers of Gaussian components on LJSpeech and LibriTTS.}
\label{fig:loglike_curve}
\end{figure*}

\subsection{Prosody predictor}
The GMM-MDN in this work is the prosody predictor whose detailed architecture is demonstrated in Figure \ref{predictor}.
The hidden state $\mathbf{h}$ of the input phoneme sequence is passed through 2 layers of 1D convolution, each followed by a ReLU, layer normalization and dropout layer. The output of the above modules is then concatenated with the previous prosody embedding and sent to a GRU. The GRU is designed to condition the prediction of the current prosody distribution on the previous prosodies. Then we project the GRU output to obtain $w_{k,i}$, $m_{k,i}$ and $v_{k,i}$, which is then transformed to the GMM parameters according to Equation (\ref{eq:softmax}) - (\ref{eq:var}).

Equation (\ref{eq:mdn_loss}) formulates the training criterion for an MDN, which is the negative log-likelihood of the observations. Here, the observations are the prosody embeddings $\mathbf{e}$, so we obtain the loss function for training the prosody predictor
\begin{equation}
\begin{aligned}
\label{eq:pp_loss}
\mathcal{L}_{\texttt{PP}} & = \sum_{k=1}^{K} -\log p\left(\mathbf{e}_k;\mathbf{e}_{<k},\mathbf{h}\right) \\
& = \sum_{k=1}^{K} -\log \left(\sum_{i=1}^{M} w_{k,i} \cdot \mathcal{N}\left(\mathbf{e}_k ; \mu_{k,i}, \sigma_{k,i}^{2}\right)\right)
\end{aligned}
\end{equation} 
where $w_{k,i}$, $\mu_{k,i}$ and $\sigma_{k,i}$ are the Gaussian parameters of the $i$-th component. The parameters are predicted given $\mathbf{h}$ and $\mathbf{e}_{<k}$.

\subsection{Training criterion}

The prosody extractor and the prosody predictor are both jointly trained with the FastSpeech2 architecture. The overall architecture is optimized with the loss function
\begin{equation}
\begin{aligned}
\label{eq:overall_loss}
\mathcal{L} & =  \beta \cdot \mathcal{L}_{\texttt{PP}} + \mathcal{L}_{\texttt{FastSpeech2}} \\
& = \beta \cdot  \mathcal{L}_{\texttt{PP}} + \left( \mathcal{L}_{\texttt{MEL}} + \mathcal{L}_{\texttt{VAR}} \right)
\end{aligned}
\end{equation}
where $\mathcal{L}_{\texttt{PP}}$ is defined in Equation (\ref{eq:pp_loss}), $\mathcal{L}_{\texttt{FastSpeech2}}$ is the loss function of FastSpeech2 which is the sum of variance prediction loss $\mathcal{L}_{\texttt{VAR}}$ and mel-spectrogram reconstruction loss $\mathcal{L}_{\texttt{MEL}}$ as described in \cite{fastspeech2}, and $\beta$ is the relative weight between the two terms.
It should be noted that we use a stop gradient operation on $\mathbf{e}$ in calculating the $\mathcal{L}_{\texttt{PP}}$, so the prosody extractor is not optimized with $\mathcal{L}_{\texttt{PP}}$ directly.

\section{Experiments and Results}
\label{sec:setup}

\subsection{Experimental setup}

LJSpeech \cite{ljspeech} is an English dataset, containing about 24 hours speech recorded by a female speaker. We randomly leave out 250 utterances for testing.

 All the speech data in this work is resampled to 16kHz for simplicity. The mel-spectrograms are extracted with 50ms window, 12.5ms frame shift, 1024 FFT points and 320 mel-bins. Before training TTS, we compute the phoneme alignment of the training data with an HMM-GMM ASR model trained on Librispeech \cite{librispeech}, and then extract the duration of each phoneme from the alignment for TTS training.

 The $\beta$ in Equation (\ref{eq:overall_loss}) is set to 0.02. An Adam optimizer \cite{adam} is used for TTS training in conjunction with a noam learning rate scheduler \cite{attalluneed}. 
The output mel-spectrogram is converted to the waveform with a MelGAN \cite{melgan} vocoder, which is trained on the same training set.

In addition to the proposed system PLP-GMM, we also build two other systems ULP and PLP-SG with different prosody modelling methods. 
1) ULP: A recent popular approach for utterance-level prosody (ULP) modelling is conditioning the synthesis on a latent variable from VAE \cite{tacotron_vae}. The dimensionality of the latent variable is set to 128. In the training stage, the latent variable is sampled from the posterior for each utterance. In the inference stage, the latent variable is sampled from the prior distribution, which is a standard Gaussian $\mathcal{N}(0, I)$. 
2) PLP-SG: For precise prosody modelling, we apply phone-level prosodies (PLP) in this work. One basic method of PLP modelling is using a single Gaussian \cite{phone_level_prosody_amazon,phone_level_3dim_quantize}. 
3) PLP-GMM: As is described in Section \ref{sec:pl_modelling}, the proposed system models the phone-level prosodies with a GMM based MDN, which means the output of the prosody predictor is the parameters of GMMs. The other configurations of PLP-GMM are the same as PLP-SG.

\subsection{The necessity of using phone-level prosodies}
\label{sec:pl_vs_ul}
 Firstly, we verify whether using the extracted ground-truth phone-level prosodies $\mathbf{e}$ is better than using utterance-level prosodies in reconstruction. We reconstruct the test set with PLP-GMM and ULP, which is guided by the extracted phone-level prosody $\mathbf{e}$ and by the utterance-level prosody sampled from the latent posterior respectively.

 \begin{table}[htbp]
\caption{Reconstruction performance (MCD) on the test set.}
\label{tab:reconstruct}
\centering
\begin{tabular}{c|c}
\hline
\textbf{Prosody Condition} & \textbf{MCD} \\ \hline
Utterance-level    & 5.22 \\ 
Phone-level    & 3.38 \\ \hline
\end{tabular}
\end{table}

Mel-cepstral distortion (MCD) \cite{mcd} is an objective measure of the distance between two sequences of mel-cepstral coefficients. We extract 25 dimensional mel-cepstral coefficients with 5ms frame shift \cite{mcd_25d} from both the synthetic speech and the corresponding ground-truth speech on the test set for computing the MCD. The results are demonstrated in Table \ref{tab:reconstruct}, where a lower MCD represents a better reconstruction performance. It can be observed that using the extracted phone-level prosodies achieves lower MCD than the utterance-level baselines. This is natural because phone-level prosody representations contain much more information about how each phoneme is pronounced than an utterance-level representation. Therefore, it is necessary to use phone-level prosodies in TTS systems for precise prosody controlling.

\subsection{The number of Gaussian components for phone-level prosody modelling}
In this section, we try to figure out how many Gaussian components are needed to model the distribution of the extracted phone-level prosodies $\mathbf{e}$. We plot the log-likelihood curves on both the training set and the test set with several different numbers of Gaussian components in Figure \ref{fig:loglike_curve}. 

We can find that the log-likelihood gap between the training and test set in the single Gaussian model is larger than that in the Gaussian mixture models. Moreover, the log-likelihood curves of Gaussian mixture models are much higher than the single Gaussian model in both the training and test set. Therefore, we can conclude that the distribution of phone-level prosodies is multi-modal and it should be modeled with Gaussian mixtures. 

Additionally, increasing the number of components also provides higher log-likelihood, because the more components enable the model to simulate more complicated distribution. However, when we double the number of the components from 10 to 20, the improvement is very limited. Therefore, we do not further increase the number of components, and use 20 components in the following GMM-based systems.

\subsection{Prosody diversity}

According to the investigation above, we train the proposed system PLP-GMM with 20 Gaussian components. In the inference stage, the Gaussian mixture distributions of phone-level prosodies are predicted, from which the phone-level prosodies $\hat{\mathbf{e}}$ are sampled. The synthesis is then guided by the sampled prosodies $\hat{\mathbf{e}}$. \footnote{Audio examples are available here \texttt{\url{https://cpdu.github.io/gmm_prosody_examples}}.}

 \begin{figure}[htbp!]
\centering
\includegraphics[width=\linewidth]{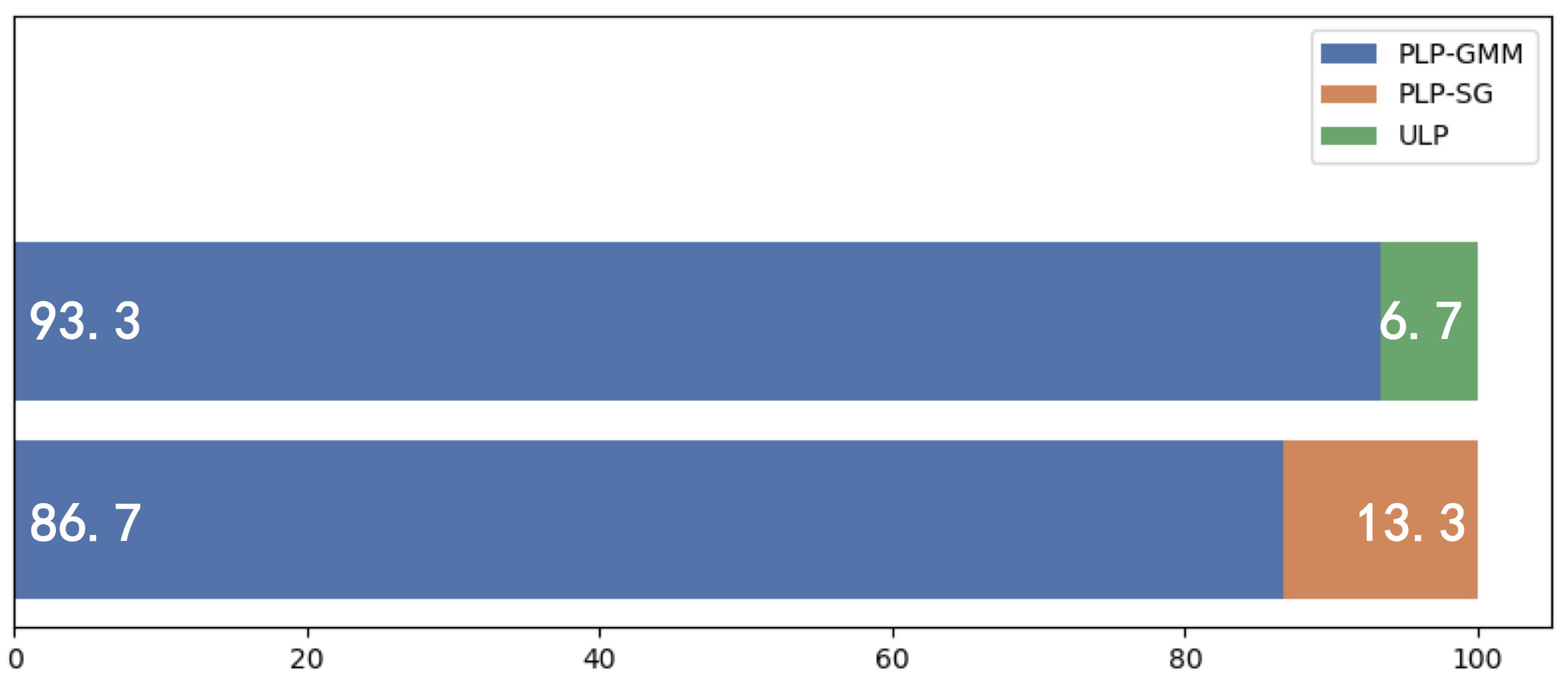}
\caption{AB preference test in terms of prosody diversity.}
\label{fig:ab_ljspeech}
\end{figure}

We compare PLP-GMM with two baseline systems ULP and PLP-SG in terms of prosody diversity.
We synthesize the utterances in the test set 3 times with various sampled prosodies. We perform AB preference tests where two groups of synthetic speech from two different TTS systems are presented and the listeners need to select the better one in terms of prosody diversity. The results are shown in Figure \ref{fig:ab_ljspeech}. We can find that PLP-GMM provides better prosody diversity in the synthetic speech than both ULP and PLP-SG.
This can be easily explained by the fact that a sequence of phone-level embeddings depicts the prosody more precisely than an utterance-level embedding and the fact that Gaussian mixtures can better model the phone-level prosody embeddings than a single Gaussian.

\subsection{Naturalness}
\begin{figure}[htbp!]
\centering
\includegraphics[width=0.9\linewidth]{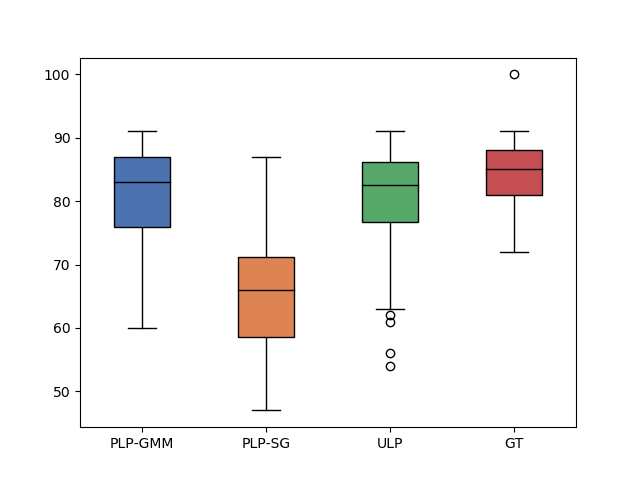}
\caption{MUSHRA test in terms of naturalness.}
\label{fig:nt}
\end{figure}

We also evaluate the naturalness of the above systems with a MUSHRA test, in which the listeners are asked to rate each utterance on a scale of 0 to 100. The speech converted back from the ground-truth mel-spectrogram with the vocoder is also rated in the test, which is denoted as GT. The results are reported in Figure \ref{fig:nt}. It can be observed that PLP-GMM synthesizes speech with higher naturalness score compared with PLP-SG because of the better phone-level prosody modelling. We can also find that ULP generates speech with similar naturalness as PLP-GMM, which can also be easily explained. In ULP, no phone-level prosody is explicitly considered. Hence, the synthetic speech contains an averaged phone-level prosodies, whose naturalness is comparable to the sampled phone-level prosodies in PLP-GMM. Finally, we notice that the median of PLP-GMM is also slightly higher than ULP, but both the medians of PLP-GMM and ULP are lower than GT.

\section{Conclusions}
\label{sec: conclusion}
In this work, we propose a novel approach that uses a GMM-MDN to model the phone-level prosodies. Our experiments first prove that using the extracted phone-level prosodies can better reconstruct the speech than using utterance-level prosodies. Then we find that the log-likelihood of phone-level prosodies increases when more Gaussian components are used, indicating that GMM-MDN can generate more natural and smooth prosody pattern than a single Gaussian. The subjective evaluations suggest that our method not only achieves a better naturalness, but also significantly improves the prosody diversity in synthetic speech without the need of manual control.



\bibliographystyle{IEEEtran}

\bibliography{mybib}

\end{document}